\documentclass[proceedings,opts]{JHEP}
\usepackage{epsfig}

\def\beq{\begin{equation}}
\def\eeq{\end{equation}}
\def\bea{\begin{eqnarray}}
\def\beaa{\begin{eqnarray*}}
\def\eea{\end{eqnarray}}
\def\eeaa{\end{eqnarray*}}
\def\bq{\begin{quote}}
\def\eq{\end{quote}}
\def\gappeq{\mathrel{\rlap {\raise.5ex\hbox{$>$}}
{\lower.5ex\hbox{$\sim$}}}}

\def\11{\mbox{l\hspace{-.52em}1}}

\def\lappeq{\mathrel{\rlap{\raise.5ex\hbox{$<$}}
{\lower.5ex\hbox{$\sim$}}}}

\conference{Heavy Flavours 8}
\title{FUTURE OF HEAVY-FLAVOUR PHYSICS}
\author{John ELLIS \\ Theoretical Physics Division, CERN \\
CH - 1211 Geneva 23 \\}
\abstract{The prospects for the coming Golden Age of heavy-flavour
physics are discussed from the perspective of one who hopes it may
provide a window onto physics beyond the Standard Model. Precise QCD
calculations are necessary for accurate determinations of CKM parameters,
and may be provided by the lattice and new perturbative techniques.
The future of CKMology in the wake of the relatively small branching
ratio for $B_d \rightarrow \pi^+ \pi^-$ are discussed, including
alternative strategies for measuring $\alpha$ and $\gamma$. The
opportunities
for measurements of rare $B$ decays are reviewed briefly, as are
possible $K$-decay windows on physics beyond the Standard Model,
in the wake of the establishment of a relatively large value for
$\epsilon'/\epsilon$. Finally, relations to other aspects of flavour
physics are discussed, including the grand unification of quark and lepton
masses in the context of neutrino-mass models motivated by the
recent oscillation data from Super-Kamiokande and elsewhere. \\
$\phantom{xxxxxxxxxxxxxxxxxxxxxxxxxxxxxx}$ \\
$\phantom{xxxxxxxxxxxxxxxxxxxxx}$
CERN-TH/99-318 $\phantom{xxxx}$ hep-ph/9910404
\\
}
\dedicated{}

\begin{document}

\section{Why Study Heavy Flavours ?}

There are many honourable answers to this question: the subject has many
challenging problems, perhaps it was your thesis topic, maybe you are in love
with the Standard Model, or want to answer Rabi's question about the muon: who
ordered that? My personal perspective is based on the hope that heavy
flavours may provide a welcome window on physics beyond the Standard Model.
However, even
if this window is closed, precision measurements of Standard Model
parameters may provide welcome indirect clues, just as the measurement of
$\sin^2\theta_W$ at LEP provided a hint of supersymmetric grand
unification. But one
should not exclude {\it a priori} the
possibility that heavy flavours may finally
extract us directly from our Standard Model straitjacket.

Heavy flavours are good places to look, because many new physics effects are
suppressed by factors $\propto (E/M)^n$, where $E$ is the energy and $M$ is
some
large mass scale. Heavy flavours $Q$ provide ready-made high energies $E\sim
m_Q$, so any new physics effects may be enhanced by comparison with
light-flavour
physics. The prime example is top physics~\cite{Simmons}: in some models
of
flavour $m_t/M =
0(1)$, and certainly the top quark Yukawa coupling $\lambda_t = 0(1)$,
so here may be the key that unlocks the door of flavour physics. As another
example, we shall see later that in a class of supersymmetric GUT models
$B(\tau\rightarrow\mu\gamma)\gg B(\mu\rightarrow e\gamma)$.

From this perspective, we need to understand QCD in order to do CKMology
better,
in the  hope of finding some signature of physics
beyond the Standard Model. This
might either appear in the form of a (relatively) big contribution to a rare
decay mode, e.g., perhaps $B(B_d\rightarrow \mu^+\mu^-) > 2 \times
10^{-10}$ (the
Standard Model prediction), or of a (relatively) small contribution to a
common
decay, e.g., perhaps the unitary triangle will turn out to be a
quadrilateral?

In any case, the golden age of heavy flavours is surely now dawning. The era of
precision electroweak physics is drawing to a close: the SLC has been
terminated
and LEP will cease operation in 2000. At the same time, new opportunities for
precision flavour physics are opening up: BaBar, BELLE, CLEO III,
CDF/D$\phi$ at
RUN II, etc.

\section{QCD Challenges}

Even though $1 \gg \alpha_s (m_Q)$ in heavy-quark physics, the crucial
challenge here is: how to calculate accurately in a theory whose
perturbative expansion parameter $\alpha_s (m_Q) \gg\alpha_{em}$ ?  The
traditional approximation scheme for any observable is to write it in the form
\beq
\sum_n~C_n (\alpha_s(\mu ) , {\Lambda\over\mu}) < O_n(\mu)>
\label{one}
\eeq
where the $C_n$ are coefficient functions, $\mu$ a factorization scale, and the
$<O_n>$  operator matrix elements or condensates to be calculated
non-perturbatively, e.g., using lattice techniques. Unfortunately, because of
weakness and/or stupidity, we can only calculate to some finite order in
$\alpha_s$. However, we know that the perturbation series is at best
asymptotic, and
there are in general renormalons. The scheme dependence they introduce can be
absorbed into higher-twist terms, but we can only calculate a few of these,
e.g.,
in heavy-quark effective theory (HQET)~\cite{Rothstein}. Not only are
perturbative expansions
asymptotic, but there may be effects that are invisible in any approximation
scheme. As discussed here~\cite{Shifman}, there may be singularities away
from the
light-cone, at $|x^2| \not= 0$, even $|x^2| \rightarrow\infty$. There are
examples such as instantons and the 't Hooft
model of such breakdowns of duality,
but the practical importance of such effects is not clear.

Some of the primary QCD puzzles in heavy-flavour physics are provided by
production
cross sections in hadron-hadron collisions. The Tevatron cross
sections
for
$J/\psi$ and $\psi^\prime$ production~\cite{Sumorok} are much larger than
expected in a na\"\i
ve colour-singlet model, leading to the proposal of colour-octet
dominance~\cite{Braaten,Rothstein}. This
reproduces well the large-$p_T$ dependence, but the overall normalization
is not
well determined theoretically~\cite{Nason}. Since the colour-octet
mechanism invokes soft
E1 gluons that do not flip spin, it predicts large
polarization~\cite{Rothstein}. However, this
has not been seen~\cite{Seth,Sumorok}, either in charmonium or bottomonium
production.

Naked heavy-flavour production is notoriously difficult to
calculate~\cite{Nason}. Near
threshold, there are large
$[\alpha_s \ln^2 (1-4 m^2_Q/\hat s)]^n$ corrections to
be resummed, and at high energies there are large
$[\alpha_s \ln (\hat s/ 4m^2_Q/)]^n$ effects. These provide potentially large
enhancements, with $O(\Lambda/m_Q)$ ambiguities at threshold. The cross
section for
$\bar bb$ production measured at the Tevatron collider~\cite{Sumorok} is
about 2.5 times higher
than the best theoretical prediction available. At HERA, the status of the
theoretical predictions is particularly confusing~\cite{Sefkow}:
$\sigma^c_{exp}/\sigma^c_{theory} \sim 1$, whilst
$\sigma^b_{exp}/\sigma^b_{theory} \sim 4$! Of particular interest to HERA-B is
the cross-section they should expect. Unfortunately, it is currently not
possible
for theory to provide precise guidance~\cite{Nason}: estimates lie in the
range
\beq
\sigma^b_{\rm HERA-B} \sim 7.6~{\rm to}~ 45~{\rm mb}
\label{two}
\eeq
with significant uncertainties associated with the choice of $m_b, \mu$ and
proton structure functions.

Heavy quarkonia offer interesting opportunities to determine $\alpha_s$, if the
systematic errors in the required lattice calculations can be controlled to the
level needed. The NRQCD value from the $\chi_c - J/\psi$ system
is~\cite{Rothstein}
$$
\alpha_s^{\overline{MS}}(M_Z) = 0.1174 (15)~(19)~(\ldots)
\eqno{(2.3a)}
$$
whilst that obtained from the $\psi^\prime$ and $J/\psi$ is
$$
\alpha_s^{\overline{MS}}(M_Z) = 0.1173 (21)~(18)~(\ldots)
\eqno{(2.3b)}
$$
where in each case the last parenthesis $(\ldots)$ is to accommodate the
lattice
errors associated with the extrapolations needed in $m_{u,d,s}$ and
sending the lattice
size $a\rightarrow 0$. The values (2.3) are to be compared with
\addtocounter{equation}{1}
\bea
\alpha_s^{\overline{MS}}(M_Z) &=& 0.119 \pm 0.002~~({\rm
LEP}@Z~\cite{LEPEWWG})~, \nonumber \\
&&0.120 \pm
0.003~(\tau~{\rm decay}~\cite{Davier})
\label{four}
\eea
The extension of NRQCD to the $\bar tt $ system can benefit from the
extremely large
mass of the $t$ quark and its relatively short lifetime, as in the
potential NRQCD (PNRQCD) approach~\cite{Beneke}.

The numbers (2.3), (\ref{four}) are of interest for testing supersymmetric GUT
predictions. It is also possible to extract from $\tau$
decays~\cite{Davier,Pich}
\beq
\overline{m_s}~(1 ~{\rm GeV}) = 234^{+61}_{-76}~{\rm MeV}
\label{five}
\eeq
which is to be confronted with the lattice value~\cite{Martinelli}
\beq
\overline{m_s}~(1 ~{\rm GeV}) = 130~{\rm to}~ 180~{\rm MeV}
\label{six}
\eeq
In addition to its great importance for QCD and CKMology, the value of
$\overline{m_s}$  is also of interest for testing models of flavour.
Bottomonium spectroscopy can  be used to estimate
$m^{\overline{MS}}_b(m_b^{\overline{MS}})$. Here, a significant advance has
recently been made with the calculation of the two-loop lattice matching
factor,
which removes an ambiguity of several hundred MeV, resulting in the
value~\cite{Martinelli}
\beq
m^{\overline{MS}}_b(m_b^{\overline{MS}}) = 4.3 \pm 0.1~{\rm GeV}
\label{seven}
\eeq
A precise value for $m_b$ is also of interest for testing the
predictions of GUTs~\cite{CEG}
and models of flavour, as discussed later.

Other aspects of heavy-hadron spectroscopy discussed here, such as
hybrids~\cite{Michael} and
excited $D^{**}$ and $B^{**}$ states~\cite{Ciulli}, have intrinsic QCD
interest, but are not so
important in a larger context.

Turning now to heavy-flavour decay matrix elements, it seems that lattice
calculations are now taking over as the most precise for many applications.
Compare, for example, the QCD sum-rule numbers for decay
constants~\cite{Braun}:
\bea
f_B &=& 160 \pm 30~{\rm MeV}~, \quad f_{B_s/f_B} = 1.16 \pm 0.09 \nonumber \\
f_D &=& 190 \pm 30~{\rm MeV}~, \quad f_{D_s/f_D} = 1.19 \pm 0.08 \nonumber \\
\label{eight}
\eea
which are `not improvable'~\cite{Braun}, with the unquenched lattice
results~\cite{Hashimoto}:
\bea
f_B &=& 210 \pm 20~{\rm MeV}~, \quad f_{B_s/f_B} = 1.16 \pm 0.04 \nonumber \\
f_D &=& 211 \pm 30~{\rm MeV}~, \quad f_{D_s/f_D} = 1.11 \pm 0.04 \nonumber \\
\label{nine}
\eea
Also impressive are the recent lattice results for the $B$
parameters~\cite{Hashimoto}:
\beq
B_B(m_b) = 0.80 \pm 0.10~, \quad B_{B_s}/B_{B_d} = 1.00 \pm 0.03
\label{ten}
\eeq
Also of direct experimental interest are the estimates~\cite{Hashimoto}
\beq
\left({\Delta\Gamma\over\Gamma}\right)_{B_S} = 0.16~(3)~(4)~, \quad
{\tau_{B^-}\over \tau_{B^0}} = 1.03 \pm 0.2 \pm 0.3
\label{eleven}
\eeq
The ratio $\tau_{\Lambda_b}/\tau_{B^0}$ could probably benefit from
further analysis.

The lattice calculations of heavy-hadron
semi-leptonic-decay form factors are now
also very competitive~\cite{Hashimoto,Lellouch}. They can be calculated
directly in
the physical region of
$q^2$, obviating the need for any extrapolation to extract CKM matrix elements.

The interpretation of non-leptonic $B$ decays has been beset by greater
theoretical uncertainties, but there have recently been promising
developments in
the formulation of the problem~\cite{Benekeetal}. A typical non-leptonic
decay
amplitude may be
written schematically as
\beq
A = {G_F\over \sqrt{2}} \sum_n V_n^{CKM} C_n(\mu ) <F|Q_n(\mu )|B>
\label{twelve}
\eeq
where our (my) primary objective is the extraction of the different CKM matrix
elements $V_n^{CKM}$, and the essence of the problem is that the hadronic
matrix
elements $<F|Q_n(\mu )|B>$ are often not suitable for lattice evaluations.
It is
an old idea that these might factorize, e.g., in two-body decays:
\beq
<M_1M_2|Q|B> ~\simeq ~<M_1|J_1|B>~<M_2|J_2|0>
\label{thirteen}
\eeq
but several issues have been outstanding, e.g., the scheme and scale
invariance,
possible non-factorizable terms, final-state-interaction (FSI) phases, etc. A
systematic extension to non-leptonic decays of the QCD factorization known in
hadronic processes~\cite{oldfactn} has recently been
made~\cite{Benekeetal}, as illustrated in Fig. 1, where there
are two distinct contributions,
distinguished by the hardness of the interactions
with the spectator quark in the $B$ meson. In Fig. 1a, soft spectator
interactions are subsumed in an exclusive form factor, there is a
short-distance
hadronic wave-function factor and a $2 \rightarrow 2$ hard-scattering
kernel $T_I$.
In Fig. 1b, the spectator interactions are hard, and the factorization is into
three short-distance wave functions and a $2\rightarrow 4$ hard-scattering
kernel
$T_{II}$. This approach formalizes perturbative QCD factorization and FSI, but
does not include power corrections. Some of these could be important,
particularly if they contain large chiral factors:
\beq
\left({\Lambda_{QCD}\over m_b} \right) \times {m^2_\pi\over
\Lambda_{QCD}(m_u+m_d)}
\simeq {1\over 2}
\label{fourteen}
\eeq
is not very small!

\EPSFIGURE[ht]
{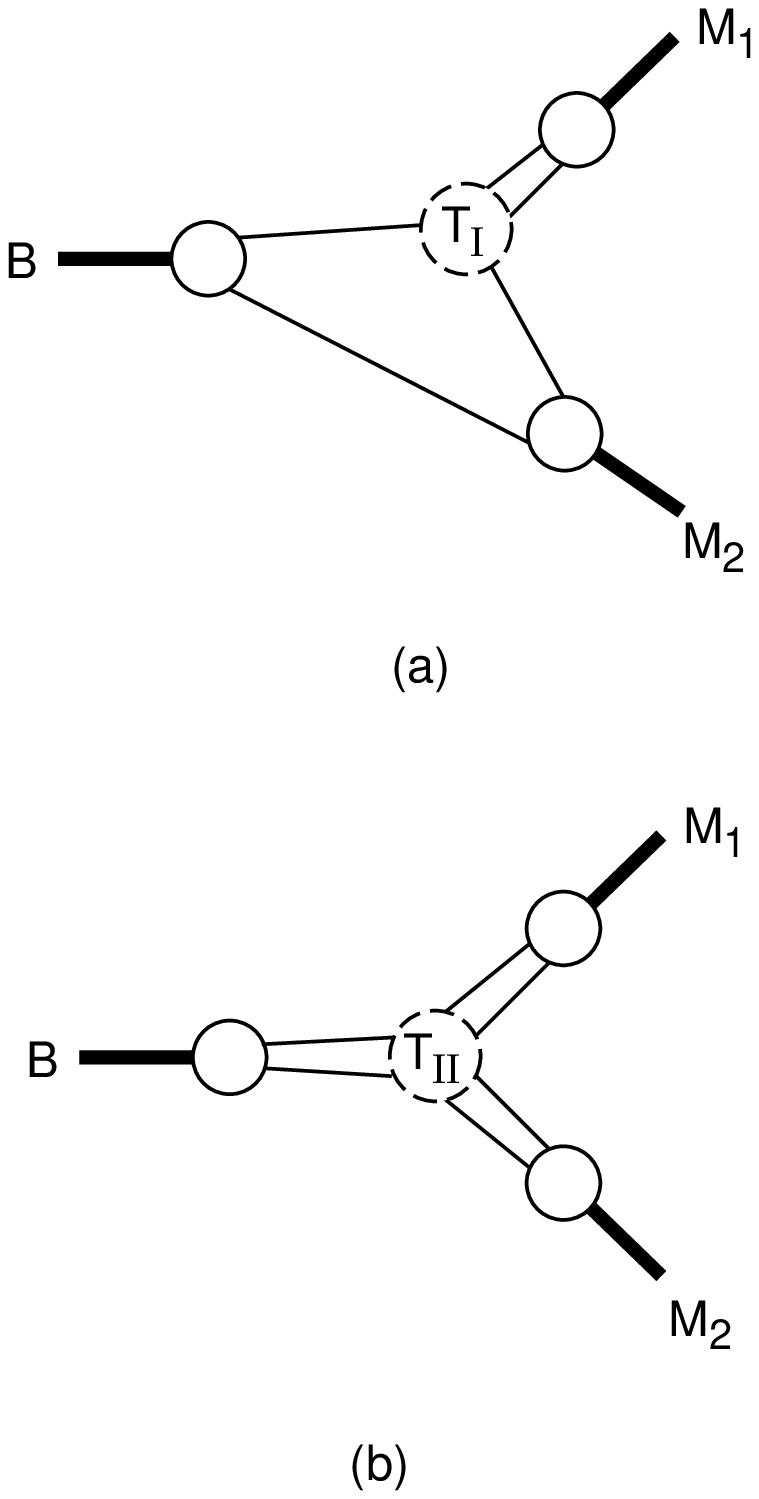,width=4.5cm}
%Fig. 1 goes here.
%\caption{\it
{Contributions to the decay of a $B$ meson into two light
mesons $M_{1,2}$, that may be factorized (a) with a $2 \rightarrow 2$
hard-scattering kernel $T_I$,
and (b) with a $2\rightarrow 4$ hard-scattering kernel $T_{ll}$
[20].}

A useful phenomenological complement to this analysis is provided by
diagramatic approaches based on Wick
rotations~\cite{Ciuchini,Silvestrini}, as
illustrated in Fig. 2, which can
be improved using the renormalization group to become scale and scheme
independent. In the honour of our Southampton hosts, I propose that Fig.
2a be
baptized a `dolphin' diagram \footnote{See~\cite{oldShifman} for an
unexpurgated
account of the baptism of `penguin' diagrams~\cite{EGNR}.}. Note that
Fig. 2b includes
rescattering, as deconstructed in Fig. 2c.

\EPSFIGURE[ht]
{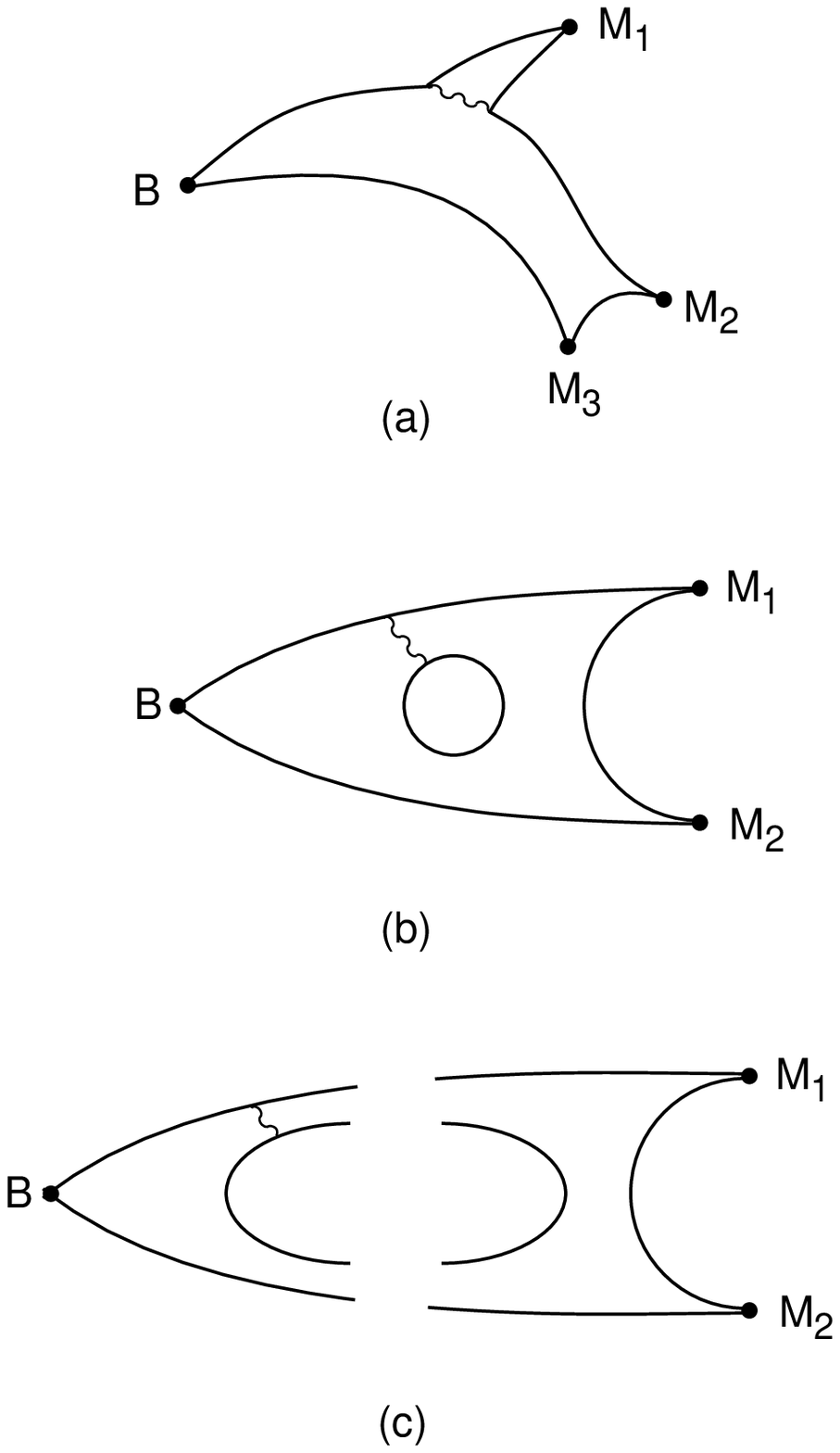,width=4.5cm}
%Fig. 2 here.
%\caption{\it
{Examples of the renormalization-group-invariant classification
of diagrams for $B$ decays~\cite{Silvestrini}, including (a) a `dolphin'
diagram for $B \rightarrow M_1 + M_2 +M_3$ decay, and (b) $B \rightarrow
M_1 + M_2$ decay, which may (c) be deconstructed to exhibit rescattering.
The wavy lines denote generic-four-quark operators.}
%}

\section{CKMology}

Pride of place in this discussion goes to the unitarity triangle
discussed here in more detail in~\cite{Plaszczynski}, of which I now
discuss various aspects in turn~\cite{Gronau}.

\noindent
\underline{$\sin 2\beta$} : The theory of the CP-violating asymmetry in
$B^0\rightarrow J/\psi~K_S$ decay is gold-plated, since the penguin
contributions are expected to be very small, and FSI are
unimportant~\cite{Dunietz}. This
decay is also
experimentally gold-plated, and the asymmetry
has already been `measured' by CDF~\cite{Bortoletto}:
\beq
\sin 2\beta = 0.79 (39)~(16)
\label{fifteen}
\eeq
We will soon have $B$-factory measurements with a precision expected to
attain
$\pm$ 0.12 to 0.05~\cite{Waldi,Suzuki}, and eventually measurements at
hadron machines
may attain a precision $\pm$ 0.01. It should also be borne in mind that the
available statistics could be doubled
by including additional decay modes such as
$B^0\rightarrow J/\psi K_L$, $\psi^\prime K_S$, etc.
Measuring $\sin 2\beta$
does not determine $\beta$ uniquely, but the ambiguity could be removed by
complementary measurements, e.g., comparing $B_d\rightarrow J/\psi K^*$ and
$B_s\rightarrow J/\psi \phi$ can fix $\cos 2\beta$, and the cascade decay
$B_d\rightarrow J/\psi (K^0\rightarrow \pi \ell\nu )$ can be used to measure
$\cos 2\beta \sin (\Delta m_K t_K)$~\cite{Dunietz}.

That was the good news $\ldots$ and now for the bad news $\ldots$

\noindent
\underline{$\sin 2\alpha$} : The prime candidate for measuring this
quantity was
$B^0\rightarrow \pi^+\pi^-$, which was unseen by CLEO until just
recently. The
good news is that this mode has now been observed~\cite{Kubota}:
\beq
B(B^0\rightarrow \pi^+\pi^-) = (0.47^{+.18}_{-.15} \pm .13) \times 10^{-5}
\label{sixteen}
\eeq
but the bad news is that this is a factor $\sim 4$ below
\beq
B(B^0\rightarrow K^+\pi^-) = (1.88^{+.28}_{-.26} \pm .06) \times 10^{-5}
\label{seventeen}
\eeq
The small ratio (\ref{sixteen}), (\ref{seventeen}) means that there must be
considerable penguin pollution, which introduces ambiguity in the determination
of $\sin 2\alpha$~\cite{Charles}, since penguins and trees have different
phases. The
improvement in theoretical calculations possible using
isospin~\cite{GL} or U-spin relations~\cite{Uspin}, or the new approach to
amplitude factorization~\cite{Benekeetal} may enable some of the penguin
pollution to be cleaned
up, but only a target error $\delta\alpha = \pm 10^0$ may be realistic.
Under
these circumstances, alternative ways to measure $\sin 2\alpha$ become more
interesting. One suggestion~\cite{Snyder} is the $B\rightarrow\rho\pi$
Dalitz plot, but here
there are questions concerning the attainable statistics and the sensitivity in
the presence of background, and another is to use $B\rightarrow D^{*\pm}
\pi^\mp$
decays, which looks tough~\cite{Dunietz}.

\noindent
\underline{$\gamma$} : In view of the above imbroglio, in particular, there is
increased interest in constraining and/or measuring $\gamma$. One line of
attack~\cite{Neubert}
is to use $B^\pm\rightarrow\pi^\pm K^0$, $\pi^0K^\pm$ and $\pi^\pm\pi^0$ decays
which receive contributions from penguins, electroweak penguins and tree
diagrams
(whose phase is $\gamma$). Using
\beq
R_* \equiv {B(B^\pm\rightarrow\pi^\pm K^0)\over 2B(B^\pm\rightarrow\pi^0
K^\pm)}
= 0.47 \pm 0.24
\label{eighteen}
\eeq
and $\bar\epsilon_{3/2} = 0.24 \pm 0.06, \delta_{EW} = 0.64 \pm 0.15$, in the
inequality
\beq
R \equiv {1-\sqrt{R_*}\over \bar\epsilon_{3/2}} \leq |\delta_{EW} - \cos
\gamma |
+ \ldots
\label{nineteen}
\eeq
where the dots denote small corrections, one finds the lower limit $\gamma >
71^0$. It is possible, in principle, to extract $\gamma$ from the following
combination of CP-violating asymmetries:
\bea
\tilde A &\equiv& {A_{CP}(\pi^0 K^\pm)\over R_*} - A_{CP} (\pi^\pm K^0)
\nonumber \\
&=& 2
\bar\epsilon_{3/2} \sin \gamma \sin \phi +  \ldots
\label{twenty}
\eea
where $\phi$ is a strong-interaction phase: by measuring both $\tilde A$
and
$R_*$, both $\gamma$ and $\phi$ can be determined~\cite{Neubert}.

Other channels offering prospects for measuring $\gamma$
include~\cite{Dunietz} $B^\pm
\rightarrow {\buildrel{(-)}\over{ D^0}} K^\pm$, ${\buildrel{(-)}\over{ B^0}}
\rightarrow D^0 K^{*0}$,
${\buildrel{(-)}\over{ B_S}}\rightarrow D^\pm_s K^\mp$~\cite{ADK} and
${\buildrel{(-)}\over{ B_0}}\rightarrow D^{*\pm}\pi^\mp$~\cite{Fleischer},
all of which may
fairly be described as `challenging'. LHCb has studied, in particular,
the
second of these channels, and concludes that it could reach a precision of $\pm
10^0$ in $\gamma$. Hadron machines, such as the LHC, are clearly the
`Promised Land' for $B_s$ physics, which will surely flow with plenty of
CP-violating `milk and honey'~\cite{Harnew}.

\section{Rare $B$ Decays}

These offer good opportunities to measure Standard Model
parameters~\cite{Greub}, and may
also be able to open windows on physics beyond the Standard
Model~\cite{Masiero}.

As an example, \underline{$b\rightarrow s\gamma$} is related to $V_{st}$,
and is
also sensitive to $M_{H^\pm} $ and $(m_{\tilde t}, m_{\chi^\pm})$ in
supersymmetric
extensions of the Standard Model. To make a precise calculation, one must
resum the
large QCD logarithms: $(\alpha_s/\pi)^{n+m}[\ln (m^2_b/M^2)]^m$. In
the
Standard
Model, the three-loop anomalous dimension and the two-loop matching conditions
are known~\cite{twoloop}, so a relatively precise prediction can be made:
\beq
B(b\rightarrow s\gamma ) = (3.32 \pm 0.14 \pm 0.26) \times 10^{-4}\phantom{\pm
0.2xx6}
\label{twentyone}
\eeq
This is to be compared with the experimental value:
\beq
B(b\rightarrow s\gamma ) = (3.15 \pm 0.35 \pm 0.32 \pm 0.26) \times 10^{-4}
\label{twentytwo}
\eeq
There is some concern about the theoretical prediction of the $E_\gamma$
spectrum, and it might in principle be preferable to compare theory and
experiment in a restricted and more reliable range~\cite{otherNeubert},
but there is certainly no
hint yet of any need for physics beyond the Standard Model.

A NLO QCD calculation is available also in the MSSM~\cite{MSSM}, in the
limit
\bea
\mu_{\tilde g} &\equiv& {\cal O} (m_{\tilde g}, m_{\tilde q}, m_{\tilde
t_1}) \gg
\mu_{\tilde w} \nonumber \\
 &\equiv& {\cal O} (m_W, m_{H^\pm}, m_{\chi^\pm}, m_{\tilde t_2})
\label{twentythree}
\eea
The comparison between (\ref{twentyone}), (\ref{twentytwo}) is an important
constraint on the MSSM parameter space, as seen in Fig. 3~\cite{EFGOSS},
where the NLO QCD calculation is used to exclude a region of parameter
space that would otherwise be permitted for Higgsino cold dark matter.
However, the available LO
calculations are of limited accuracy when the simplified limit
(\ref{twentythree}) is applicable, so it would be good to generalize the
NLO QCD
calculation beyond the region (\ref{twentythree}) of MSSM parameter space.

\EPSFIGURE[ht]
{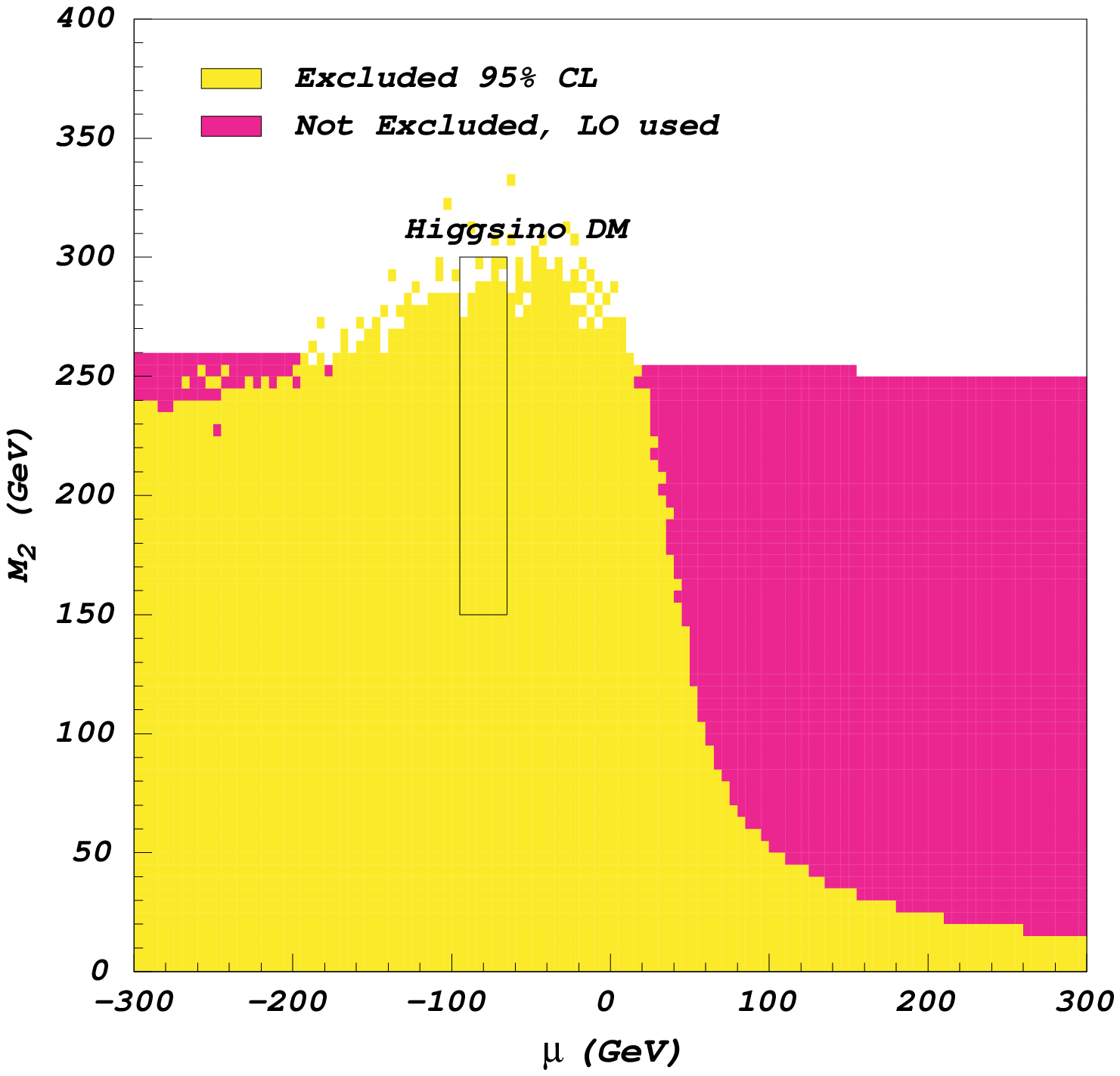,width=6cm}
%\caption{\it
{Regions of the $\mu, M_2$ plane of the MSSM which are
allowed
for Higgsino dark matter (rectangular box) but disallowed by the $b
\rightarrow s \gamma$ analysis at NLO (light shaded region): the dark
shading indicates a region allowed by a LO analysis. The following values
of the other MSSM parameters have
been chosen for this illustration:
$m_0 = 1000$ GeV, $m_A = 500$ GeV and tan$\beta = 2.5$~\cite{EFGOSS}.
}

The related decay \underline{$b\rightarrow d\gamma$} is equally calculable, and
provides a way of measuring $V_{td}$. It may also provide an opportunity for
physics beyond the Standard Model, such as supersymmetry.

The decays \underline{$b\rightarrow s,d\ell^+\ell^-$} provide additional
opportunities for testing such theories, in particular via the measurement of a
CP-violating forward-backward asymmetry. Personally, I also have a soft
spot for
\underline{$b\rightarrow s,d\bar\nu\nu$} decay, though
this will be very difficult to observe.

The decay
\underline{$B_s\rightarrow \mu^+\mu^-$}  is expected to appear with a branching
ratio of $(3.5\pm 0.1)\times 10^{-9}$ in the Standard Model. For
comparison, the
present experimental upper limit is $2.6\times 10^{-6}$. On the other hand, it
should be measurable at the LHC: a CMS study indicates that as many as 26
events
could be observed. Even the decay \underline{$B_d\rightarrow \mu^+\mu^-$}
with an
expected Standard Model branching ratio of $(1.5\pm 0.9)\times 10^{-10}$
may not
lie beyond the reach of the LHC. CDF already has established an
upper limit of
$8.6\times 10^{-7}$, and CMS may be able to observe up to
four
events. These
decays are vulnerable to physics beyond the Standard
Model such as $R$ violation
in supersymmetric models, and so could provide
an interesting window on new
physics.

\section{Beyond the Standard
Model ?}

The best prospects for new physics at the TeV scale, and hence the
most
amenable to accelerator experiments, may be offered by the problem of
mass, and
hence associated with the Higgs sector and/or 
supersymmetry. In
the MSSM, all the
renormalizable couplings are related to those in the
Standard Model, including
the gauge coupling $g_a$, the Yukawa couplings
$\lambda_{ijk}$ and their
associated CKM phases. In addition, there are the
standard soft
supersymmetry-breaking parameters of the generic
forms
\beq
(m^2_0)^j_i \phi^i \phi^*_j~, \quad {1\over 2} M_a \tilde V_a
\tilde V_a~, \quad
A_{ijk} \lambda_{ijk}
\phi^i\phi^j\phi^l
\label{twentyfour}
\eeq
i.e., scalar and gaugino masses
and 
soft trilinear couplings, respectively, and a
term $B\mu \bar HH$
associated with Higgs mixing. There is no good theoretical
reason known why
the $(m^2_0)^j_i$ should diagonalize in the same basis as the
fermion
masses, and they do not do so in generic string models, for
example.
However, there are important constraints 
on the $(m^2_0)^j_i$
from flavour physics
and CP physics~\cite{EN}, to which the most obvious
solution is that they
are universal~\cite{BG}.
There are in particular two
CP-violating phases which are constrained by the
experimental upper limits
on the electric dipole moments of the electron and
neutron, $d_e$ and
$d_n$. The interpretation of $d_e$ is cleaner, since there are
several
different operators contributing to $d_n$ (e.g., involving
$s$
quarks~\cite{EF} as
well as the valence $u$ and quarks), which could
introduce cancellations, and
hence make its interpretation more
uncertain~\cite{Bartl}.

In addition to the above supersymmetric
possibilities for new physics, one
should also bear in mind the
possibilities of non-standard soft
super-symmetry-breaking
terms~\cite{JJ}:
\beq
\phi^i \phi^*_j \phi^*_k, \quad \phi^i
\phi^j
\label{twentyfive}
\eeq
where the $\phi^i$ are generic complex
chiral scalar fields, and
$R$-violating interactions~\cite{Rviol}.

The
recent confirmation by KTeV~\cite{Barker} and NA48~\cite{Lubrano} of
the
surprisingly large value of
$\epsilon^\prime/\epsilon$ found previously by
NA31~\cite{NA31,DAgostini} has
rekindled interest in
possible supersymmetric effects
on CP-violating
observables~\cite{SusyCP}. In the Standard
Model, the first
crude calculation was made in~\cite{EGN}, and has
subsequently been greatly
refined - the fact that the data agree
with the estimate
$\epsilon'/\epsilon \sim 1/450$ given in~\cite{EGN}
is pure coincidence!
One calculates nowadays~\cite{Jamin,Eeg,Hambye,Belkov,newMartinelli} 
\bea
&&{\rm
Re}\left({\epsilon^\prime\over\epsilon}\right) \simeq  13 {\rm
Im}\lambda_t
\left({130~{\rm
MeV}\over
m_s(m_c)}\right)^2\left({\Lambda^{(4)}_{\overline{MS}}\over
340~{\rm
MeV}}\right)\nonumber \\
&&\times 
\left[ B_6
(1-\Omega_{\eta\eta^\prime}) - 0.4 B_8 \left({m_t(m_t)\over 165
~{\rm
GeV}}\right)^{2.5}\right] \nonumber \\
\label{twentyseven}
\eea
where
Im $\lambda_t = (1.33 \pm 0.14)\times 10^{-4}$ is the CP-violating
CKM
factor, $B_6$ is a penguin operator matrix element factor estimated to
be $1.0
\pm 0.3$, $\Omega_{\eta\eta^\prime}$ is an electroweak penguin
correction and
$B_8$ is another matrix element factor estimated to be $0.8
\pm 0.2$. Because of
all the uncertainties~\cite{Blum} and the possibilities of
cancellations, it is difficult to
make a precise estimate of
Re$(\epsilon^\prime/\epsilon)$. Formal analyses by two
theoretical groups
have recently yielded~\cite{Jamin,newMartinelli} 
\beq
{\rm
Re}\left({\epsilon^\prime\over\epsilon}\right) = (7.7^{+6.0}_{-3.5})
\times
10^{-4}~, \quad (4.7^{+6.7}_{-5.9})\times
10^{-4}
\label{twentyeight}
\eeq
which might suggest a discrepancy with the
experimental world average:
\beq
{\rm
Re}\left({\epsilon^\prime\over\epsilon}\right) = (21.2 \pm 4.6)
\times
10^{-4}
\label{twentynine}
\eeq
However, in view of the theoretical
uncertainties, it seems safer to quote the
following envelope of
predictions~\cite{Jamin}:
\beq
1\times 10^{-4} < {\rm
Re}\left({\epsilon^\prime\over\epsilon}\right) <
28\times
10^{-4}
\label{thirty}
\eeq
which is not in significant disagreement with
experiment (\ref{twentynine}).
Nevertheless, it is clear that the measured
value of
Re$(\epsilon^\prime/\epsilon)$ tends to favour relatively large
values of
${\rm Im}\lambda_t, \Lambda^{(4)}_{\overline{MS}}$
and
particularly $B_6$~\cite{Eeg,newMartinelli}, as well as relatively
small
values of $B_8, m_s$ and $m_t$. The least one can say, within the
context
of the Standard Model, is that the experimental 
value
(\ref{twentynine}) requires
big penguins: exotic~\cite{FM} emperors,
maybe?

This being said, even incorporating the constraints
from
$\epsilon^\prime/\epsilon$ and $K_L\rightarrow\mu^+\mu^-$, there is a
window of
opportunity for a significant contribution from CP violation
beyond the Standard
Model. One possibility is an enhanced $Zs\bar d$
vertex,
which could enhance
substantially the expected branching ratio for
the rare CP-violating decay mode
$K_L\rightarrow\pi^0\bar\nu\nu$, as well
as
$K_L\rightarrow\pi^0e^+e^-$ and $K^+ \rightarrow\pi^+\bar\nu\nu$, so
it
would be interesting
to push the experimental sensitivities for all
these modes down to the Standard
Model predictions, which are
$B(K_L\rightarrow\pi^0\bar\nu\nu) =
0.4 \times 10^{-10}$,
$B(K_L\rightarrow\pi^0e^+e^-) = 0.7 \times
10^{-11}$ and
$B(K^+\rightarrow\pi^+\bar\nu\nu) = 1.1 \times 10^{-10}$,
respectively.
They could be as large as 
$B(K_L\rightarrow\pi^0\bar\nu\nu)
= 
1.2 \times
10^{-10}$, $B(K_L\rightarrow\pi^0e^+e^-) = 2.0 \times
10^{-11}$ and
$B(K^+\rightarrow\pi^+\bar\nu\nu) = 1.7 \times
10^{-10}$,
respectively~\cite{SusyCP}.

There are many related
opportunities for signatures of physics beyond the
Standard Model in $b$
decays, notably~\cite{Masiero}: squark mixing
effects in
$b\rightarrow
d\gamma$? unexpected CP violation in $b\rightarrow s\gamma$?
rate enhancements
and CP-violating asymmetries in $b\rightarrow s,d
\ell^+\ell^-$? enhanced rates
for $B_{s,d}\rightarrow \mu^+\mu^-$? Other
signatures to bear in mind are the
possibilities that $B\rightarrow J/\psi
K_S$, $\phi K_S$ and $D^0\pi^0$ might
yield different values of $\beta$,
and that the unitarity triangle could be
distorted by new contributions to
$B_s-\bar B_s$ mixing~\cite{AliLondon}.

Although they lie beyond the
limits of heavy-flavour physics, I should also like
to advertize the
physics interests of some 
related experiments. One is a proposal
for a
new-generation experiment on the neutron electric dipole
moment. Another
is
the possibility that $\mu\rightarrow e\gamma$ decay might show
up
`close' to the
present experimental upper limit, which is motivated by
the evidence for lepton
flavour violation via neutrino oscillations,
particularly in the context of
supersymmetric GUTs~\cite{EGLLN}. A related
heavy-flavour opportunity is
provided by
$\tau\rightarrow\mu/e\gamma$
decay: estimates in the same supersymmetric GUT
framework suggest the
possibility that $B(\tau\rightarrow
e/\mu\gamma)\gappeq
10^{-9}$~\cite{EGLLN}, which might be accessible to CMS
at the LHC.

\section{Relations to Other Physics ?}

We have already seen
many potential interfaces of heavy-flavour physics with
extensions of the
Standard Model related to its 
outstanding problems of mass (via
the Higgs
sector and supersymmetry) and of unification (via  GUTs). However,
the
outstanding contributions of the next generation of heavy-flavour
experiments
will surely be towards unravelling the problem of flavour,
which is the most
baffling puzzle raised by the Standard Model. It is not
that we lack clues: the
flavour sector contains at least 13 parameters (6
quark masses, 3 lepton masses
and the 4 CKM angles), most of which have
been measured to some level. Also,
theorists abound with ideas, but their
predictions are more often qualitative
than  quantitative. Perhaps the next
generation of heavy-flavour experiments will
provide more clues: it will
certainly provide more precise measurements that may
bust some of the GUT
flavour models.

GUTs can predict quark masses in terms of lepton masses,
because quarks
and leptons are
linked in common GUT multiplets. The
prototype relation was~\cite{CEG}
\beq
m_b =
m_\tau
\label{thirtyone}
\eeq
before renormalization. The effective value
of $m_b$ varies with the energy
scale, as has been verified by DELPHI at
LEP~\cite{DELPHImb}, and the
renormalization group can
be used to calculate
the physical value of
$m^{\overline{MS}}_b(m_b^{\overline{MS}})$ if one
knows the spectrum of
particles between here and $m_{GUT}$ (the Standard
Model? the MSSM?). Starting
from (\ref{thirtyone}), such a calculation
yields~\cite{CWetal}
\beq
m^{\overline{MS}}_b(m_b^{\overline{MS}}) \simeq
4.5~~{\rm to}~~5~~{\rm GeV}
\label{thirtytwo}
\eeq
in the MSSM, with the
details depending on the sparticle thresholds, the
appropriate value of
$\alpha_s$, etc.

However, the analogous relations: $m_s \leftrightarrow
m_\mu$ and 
$m_d \leftrightarrow m_e$ are unsuccessful, which is not
surprising in the
context of GUT models of flavour. These introduce
higher-order terms in the mass
matrices~\cite{EG}:
\beq
m_b \left(\matrix{
\epsilon^n & \epsilon^m & \epsilon^p \cr
 \epsilon^{m^\prime} & \epsilon^q
& \epsilon^r \cr
 \epsilon^{p^\prime} & \epsilon^{r^\prime} &
1}\right)
\leftrightarrow m_\tau
\left(\matrix{ \epsilon^{\tilde n} &
\epsilon^{\tilde m} & \epsilon^{\tilde p} \cr
 \epsilon^{\tilde p^\prime} &
\epsilon^{\tilde q} & \epsilon^{\tilde r} \cr
 \epsilon^{\tilde p^\prime} &
\epsilon^{\tilde r^\prime} & 1}\right)
\label{thirtythree}
\eeq
where
$\epsilon$ is a small parameter, and the extra terms 
(which are indicated
only in order of magnitude) may be related to
non-renormalizable
interactions and/or approximate symmetries:
$U(1)$? a
non-Abelian
group?

New light on such models may be 
cast~\cite{CELW} by models of
neutrino masses in GUTs and the
emerging indications of neutrino
oscillations. Most theoretical models of
neutrino masses are based on the
generic seesaw
mechanism~\cite{seesaw}:
\beq
(\nu_L,\nu_R)~~\left(\matrix{0&m\cr
m^T&M}\right)~
~\left(\matrix{\nu_L\cr\nu_R}\right)
\label{thirtyfour}
\eeq
where each
entry is to be understood as a $3\times 3$ matrix in flavour
space,
the
$\nu_R$ are right-handed singlet states, $m = {\cal O}(m_q,m_\ell)$ and
$M
= {\cal O}(m_{GUT})$
$\gg m_W$. After diagonalization,
(\ref{thirtyfour}) yields light neutrino masses
\beq
m_\nu = m {1\over M}
m^T
\label{thirtyfive}
\eeq
and one naturally obtains $m_\nu \sim 10^{-2}$
eV if, e.g., $m \sim$ 10 GeV and
$M \sim 10^{13}$ GeV. Thanks to the
matrices $m$ and $M$ appearing in
(\ref{thirtyfive}), one can expect
non-trivial mixing between the light-neutrino
flavour eig- enstates,
parametrized by a mixing matrix $V_\nu$, and there will in
general also be
mixing between the charged-lepton flavour eigenstates,
parametrized by
$V_\ell$. Thus we obtain a measurable neutrino
mixing
matrix~\cite{MNS}
(between the neutrino and charged-lepton mass
eigenstates) analogous to $V_{CKM}$:
\beq
V_{MNS} = V_\ell
V^\dagger_\nu
\label{thirtysix}
\eeq
The mixing observed experimentally
might arise from $V_\ell$, or $V_\nu$, or
both, and any mixing in $V_\nu$
might arise from $m$ or $M$, {\it a priori}. The
neutrino mixing angles
need not be small: one can  easily construct models in
which the powers of
$\epsilon$ in the mass matrices (\ref{thirtythree}) are
different, the same
is true {\it a fortiori} of $m$, and we have no hints about
the structure
of $M$~\cite{ELLN}.

The indications from atmospheric neutrino
data~\cite{nudata} are for
maximal $\nu_\mu -
\nu_\tau$ mixing with a
difference in mass squared $\Delta m^2\sim 3\times
10^{-3}$ eV$^2$. The
solar neutrino data indicate mixing of $\nu_e$ with
$\nu_\mu$ and/or
$\nu_\tau$, with three possible scenarios: $\Delta m^2 \gappeq
10^{-5}$
eV$^2$ and large mixing, $\Delta m^2 \sim 10^{-5}$ eV$^2$ and small
mixing,
or $\Delta m^2 \sim 10^{-10}$ eV and large mixing. The last alternative
may
be disfavoured by the electron energy spectrum measured by
Super-Kamiokande,
and the second solar scenario is being constrained by
constraints on a day-night
difference in the Super-Kamiokande data. The
first solar scenario will be probed
by the KamLAND experiment, that is
already observing reactor
neutrinos. The atmospheric
neutrino region will
be probed by long-baseline 
accelerator neutrino experiments:
K2K in Japan
and MINOS in the U.S. are intended to look for
$\nu_\mu$ disappearance,
and
the CERN-Gran Sasso project is intended primarily to search
for
$\nu_\tau$
appearance. The investment in these projects is modest compared
to that in $B$
factories and experiments, but this may change. One of the
interesting options
for the future is a $\nu$ factory based on a muon
storage
ring~\cite{muSR}, which may be able
to extend the measurements of
neutrino oscillations to CP-violating
observables~\cite{CPnu}.

What does
this (very) light-flavour physics have to do with
heavy flavours? There
are
important implications for the interpretation of quark-lepton
mass
unification relations, and hence for theories of flavour~\cite{CELW}.
One
effect is a change
in the mass renormalization at scales $\mu$:
$m_{\nu_R} < \mu < m_{GUT}$, which
we parametrize by $\xi_N$. A second
effect is that 
non-trivial diagonalization of
the lepton mass matrix
(\ref{thirtythree}) 
may alter significantly the relevant mass
eigenvalue.
For example, if
\beq
m^0_D = A \left(\matrix{C&0\cr 0&1}\right)~, \quad
m^0_E = \left(\matrix{x^2 &
x\cr x&1}\right)
\label{thirtyseven}
\eeq
one
finds after diagonalization that
\beq
{m_D\over m_E} = {1\over
1+x^2}
\label{thirtyeight}
\eeq
As a result of these two effects, the
appropriate extrapolated ratio to compare
with GUT predictions is modified
to~\cite{CELW}
\beq
{\tilde m_\tau (m_{GUT})\over m_b(m_{GUT})} =
(1+x^2)~~\sqrt{{2\over 1+\xi^2_N}}
\label{thirtynine}
\eeq
As a
consequence, GUT mass unification can be maintained for any value
of
$\tan\beta$, the ratio of MSSM Higs vev's, whereas it was often thought
possible
only for $\tan\beta$ either very large or very
small.

\section{May You Live in Interesting Times}

This used to be
considered a curse in ancient China, but may nowadays be
considered a
blessing. You are now entering what may turn out to be the Golden
Age of
heavy-flavour physics, with Babar, BELLE, CLEO III and HERA-B starting
to
take data, CDF and D$\phi$ soon returning to the fray, and LHCb and
possibly
BTeV on the horizon. In parallel with this experimental
cornucopia, many
theoretical tools are maturing, such as the lattice, HQET,
(P)NRQCD, etc. Thus,
you, the heavy-flavour community, will soon be having
plenty of fun. You will be
able to measure precisely the parameters of the
Standard Model, perhaps
revealing
hints for new physics beyond the Standard
model. If you are lucky, you may even
find direct evidence for new physics.
Happy hunting!

\end{document}